\documentclass[manuscript]{aastex}

\usepackage[section]{placeins}
\usepackage{multirow}

\begin{document}

\title{DA white dwarfs observed in LAMOST pilot survey}

\author{Yue-Yang Zhang\altaffilmark{1,2}, Li-Cai Deng\altaffilmark{1}, Chao Liu\altaffilmark{1}, Sebastien L\'epine\altaffilmark{3}, Heidi Jo Newberg\altaffilmark{4}, Jeffrey L. Carlin\altaffilmark{4}, Kenneth Carrell\altaffilmark{1}, Fan Yang\altaffilmark{1,2}, Shuang Gao\altaffilmark{1}, Yan Xu\altaffilmark{1}, Jing Li\altaffilmark{1}, Hao-Tong Zhang\altaffilmark{1}, Yong-Heng Zhao\altaffilmark{1}, A-Li Luo\altaffilmark{1}, Zhong-Rui Bai\altaffilmark{1}, Hai-Long Yuan\altaffilmark{1}, Ge Jin\altaffilmark{5} }
\affil{$^1$Key Lab for Optical Astronomy, National Astronomical Observatories, Chinese Academy of Sciences, Beijing 100012, China}
\affil{$^2$University of Chinese Academy of Sciences, Beijing 100049, China}
\affil{$^3$Department of Astrophysics, Division of Physical Sciences, American Museum of Natural History, Central Park West at 79th Street, New York, NY}
\affil{$^4$Department of Physics, Applied Physics and Astronomy, Rensselaer Polytechnic Institute, 110 8th Street, Troy, NY 12180, USA}
\affil{$^5$University of Science and Technology of China, Hefei 230026, China}

\begin{abstract}
A total of $\sim640,000$ objects from LAMOST pilot survey have been publicly released. In this work, we present a catalog of DA white dwarfs from the entire pilot survey. We outline a new algorithm for the selection of white dwarfs by fitting S\'ersic profiles to the Balmer H$\beta$, H$\gamma$ and H$\delta$ lines of the spectra, and calculating the equivalent width of the CaII K line. 2964 candidates are selected by constraining the fitting parameters and the equivalent width of CaII K line. All the spectra of candidates are visually inspected. We identify 230 (59 of them are already in Villanova and SDSS WD catalog) DA white dwarfs, 20 of which are DA white dwarfs with non-degenerate companions. In addition, 128 candidates are classified as DA white dwarf/subdwarfs, which means the classifications are ambiguous. The result is consistent with the expected DA white dwarf number estimated based on the LEGUE target selection algorithm.
\end{abstract}


\section{Introduction}

White dwarf stars are the most common remnants of stellar evolution, and provide a record of the star formation and evolution history of the Milky Way. DA white dwarfs (DAWD), the white dwarfs with pure-hydrogen atmosphere, are the most common subset. Large samples of white dwarfs are useful in many studies. The ages of Galactic populations can be determined by constraining the luminosity (Harris et al. 2006) and mass functions of white dwarfs (Kepler et al. 2007).  The process of mass loss in stellar evolution can also be studied through an initial-final mass relation of stars that end up with white dwarfs, which is related to the chemical evolution of the Galaxy (Catal\'an et al. 2008, Kalirai et al. 2008, Kalirai et al. 2009).

Fourteen years ago, McCook \& Sion (1999) presented a catalog of 2249 white dwarfs as the Villanova catalog of spectroscopically identified white dwarfs.  The current MS catalog contains 14,325 entries. The Sloan Digital Sky Survey (SDSS) and other surveys such as the Palomar-Green (PG) survey and the SNIa Progenitor surveY (SPY) have remarkably extended the number of white dwarfs in the last decade (Harris et al. 2003, Kleinman et al. 2004, Eisenstein et al. 2006 and Kleinman et al. 2013 for SDSS, Green et al., 1986 for the PG survey, Napiwotzki et al. 2001 for SPY).
Recently, Kleinman et al. (2013) presented 19712 spectroscopically identified white dwarf stars, 12843 of which are DAWDs. The white dwarfs observed so far are mainly within a few hundred pc to 1-2 kpc in the Milky Way because of their low surface brightness. They are a mixture of the thin disk, the thick disk and the galactic halo population. The distances of DAWDs can be estimated by using the multi-band synthetic photometry based on the model white dwarf spectra (Holberg \& Bergeron 2006), which are in good agreement with distances obtained from trigonometric parallaxes. Galactic extinction of the WDs can be estimated from the spectra of white dwarfs (Holberg et al. 2008); obtaining large samples of DAWD spectra is important for studying the complicated extinction environment in the disk.

LAMOST (Large sky Area Multi-Object fiber Spectroscopic Telescope; also named Guo Shou Jing Telescope) is a Chinese national scientific research facility operated by National Astronomical Observatories, Chinese Academy of Sciences (NAOC). It is a reflecting Schmidt telescope with 4000 fibers in a field of view of 20 deg$^2$ in the sky. The wavelength resolution of the low resolution grating is $R\sim1800$ with total wavelength coverage of $3700 \lesssim \lambda \lesssim 9100$~\AA, which is divided into red and blue spectroscopic channels (Cui et al. 2012, Zhao et al. 2012).

LAMOST is a powerful instrument to survey the sky spectroscopically and will obtain millions of stellar spectra (LEGUE, LAMOST Experiment for Galactic Understanding and Exploration). The LAMOST survey started in September 2012 and will run for 5 years. LEGUE will observe over 2.5 million stars in the Galactic halo and more than 5 million stars in the low galactic latitude areas to study the Galactic halo and disk components except for the GAC part (see below). The revealed structure will provide a better understanding of the star formation history, the formation and evolution of the Galaxy and the structure of the gravitational potential.

The LEGUE survey consists of three parts in terms of sky coverage, namely the Galactic spheroid, the disk and the Galactic AntiCenter (GAC) (Zhao et al. 2012, Deng et al. 2012). The spheroid covers $|b|>20^\circ$; the GAC covers $|b|\leq30^\circ$ and $150^\circ\leq l \leq 210^\circ$; and the disk part covers covers as much of the low latitude sky area ($|b|\leq20^\circ$) as is available (Deng et al. 2012). Using weather and other site conditions (Yao et al. 2012), a simulation of the number of targets per deg$^2$ was undertaken and presented in Figure 1 of Deng et al. (2012). It reveals that a large fraction of stellar spectra will be obtained in the Galactic anticenter region. It is a good opportunity to study the Galactic disk. In order to assess the instrumental performance in survey mode and the feasibility of the science plan (see Deng et al. 2012 for LEGUE science plan), the LAMOST pilot survey started from October 2011 and ended in June 2012, including nine full moon cycles (Zhao et al. 2012).

In this paper, we report spectroscopically identified DAWDs in the LAMOST pilot survey. In Section 2, the details of the LAMOST pilot data release are presented and the database for DAWD search is analyzed. The method of DAWD candidate selection is discussed in Section 3. The catalog of DAWDs is shown in Section 4.
Finally in Section 5, a summary of the DAWDs in the pilot survey is presented and some DAWDs with special features are discussed. A prediction of DAWD number in LAMOST 5-year regular survey is made. The spectral features used to identify DAWDs are described in the Appendix.

\section{Data}

\subsection{LAMOST pilot data release}

A total of $\sim640,000$ spectra in 381 plates from the LAMOST pilot survey have recently been released (Luo et al. 2012). Figure 1 shows the spatial distribution of the entire pilot data (blue dots), overlapping the footprint of pilot survey (gray areas).

The Galactic spheroid survey footprint is different for dark nights and bright nights. The sky area of dark night observations was selected to cover GD-1 tidal stream (Grillmair \& Dionatos 2006), a low Galactic latitude ``Anticenter Box'' to cover several known Milky Way substructures (Sagittarius stream, the Anticenter Stream and the Eastern Banded Structure), and to fill in the remaining fibers in the extragalactic observing area when the surface density of available galaxies and QSOs is low (see Fig 1 in Yang et al. 2012). The sky coverage of the spheroid bright survey, along with GAC area, is a contiguous stripe centered at a declination of 29$^\circ$ to probe the Galactic populations (see Fig 1 in Zhang et al. 2012). The disk survey area consists of 8 plates along the Galactic plane, covering 22 previously identified open clusters (see Fig 1 in Chen et al. 2012). Note that several plates are not within the sky area of the designed footprint; these plates were put forward after the survey began.

The dark night observations target stellar objects with $14^m \lesssim r \lesssim 19.5^m$ and the objects for bright night plates have $r \lesssim 16.5^m$. The exposure time for dark night plates is $3\times30$ minutes and for bright plates is of order $3\times10$ minutes, depending on the distribution of the brightness of the objects (Zhao et al. 2012).

The summary of plate numbers in the spheroid, GAC and the disk is presented in Table 1. Plates in each region are separated by bright or dark observations and low ($|b|<30^\circ$) or high Galactic latitudes ($|b|>30^\circ$) of the plate centers.

Figure 2 shows the distribution of signal to noise ($S/N$) in the $g$ and $r$ bands of bright (solid) and dark plates (dashed). Figure 3 shows the distribution of magnitude $g$ and $r$ and color $g-r$ of bright and dark plates. The magnitude and color distributions of the $S/N_g>10$ and $S/N_r>10$ sub-samples are also plotted in Figure 3 as dashed lines. The histograms are normalized by the total number of pilot survey spectra in the bright or dark nights, respectively. The figure shows the actual magnitude limits on bright and dark nights. There are more missing stars with $S/N_g<10$ than with $S/N_r<10$, indicating that the spectrographs are less efficient on the blue end. In the high $S/N$ sub-sample of the $g-r$ distribution, many red targets are eliminated because the $S/N$ in the $g$ band is too low.

In order to generate a sample of DAWDs from the current dataset with high efficiency (i.e., finding almost all DAWDs in a dataset), a criterion of $S/N_g>10$ is applied. This band was selected since white dwarfs are very bright on the blue end of the spectrum. This selection retained 53\% of the spectra from the full sample.

\subsection{Input catalog design of the LAMOST pilot survey}

The input catalogs of the spheroid, GAC and the disk are selected from different photometric survey databases. SDSS DR8 photometry is used for the spheroid dark survey (Yang et al. 2012). 2MASS Point Source Catalog is used for the spheroid bright survey as the bright-end complement to SDSS DR8 photometry (Zhang et al. 2012). The input catalog of GAC is from Xuyi photometric survey (Smith et al. 2012, Deng et al. 2012). The input catalog of the disk is from the PPMXL astrometric catalog which provides $B$/$R$/$I$ magnitudes from USNO-B and $J$/$H$/$K$ magnitudes from 2MASS (Chen et al. 2012).

The target selection of the three parts of LEGUE can be described by the LEGUE pilot survey target selection algorithm presented in Carlin et al. (2012). The alogorithm selects candidates based on a general probability function
\begin{equation}
P_{\rm j, D} = \frac{K_{\rm D}}{[\Psi_0(\lambda_{\rm i})]_{\rm j}^{\alpha}} f_{\rm i} (\lambda_{\rm i}),\label{eq1}
\end{equation}
\noindent where $\lambda_{\rm i}$ denotes any observable (i.e., photometry, astrometry, or any combination of observed quantities) and $\Psi_0(\lambda_{\rm i})$ is the statistical distribution function of the observable $\lambda_{\rm i}$. $K_{\rm D}$ is a normalization constant to ensure that the probabilities sum to one. The $f_{\rm i} (\lambda_{\rm i})$ term is an optional function that can be used to overemphasize targets in particular regions of parameter space. The density of stars is calculated in multidimensional observable space, and the candidates are weighted by a power of this ``local density'' (Zhang et al. 2012).

The local density of the spheroid dark and bright survey is defined in ($r$, $g-r$, $r-i$) space or ($J$, $J-H$, $J-K$) space (for stars with only 2MASS photometry in the bright survey). $\alpha$ is set to be 1/2, indicating weighting by the inverse square root of the local density. Extra overemphasis is applied to the dark survey: A linear ramp bias function in $g-r$, beginning at $g-r=1.1^m$ and increasing blueward with a slope of 2.5, and a linear ramp bias function in $r$ magnitude, beginning at $r=17.5^m$, increasing toward brighter stars with a slope of 1.0 (Carlin et al. 2012). As a result, stars with $g-r<0.0^m$, among which are most of the white dwarfs, BHBs and BSs, can be overemphasized by a factor of at least 3 (see Table 1 in Carlin et al. 2012). In both the disk and the GAC region, targets are sampled evenly in color and magnitude space, i.e. ($I$, $B-I$) space for the disk and ($r$, $g-r$, $r-i$) space for GAC (Deng et al. 2012, Chen et al. 2012). This selection is equivalent to $\alpha=1$ in Equation~(\ref{eq1}).  The rare objects are strongly overemphasized. Figure 4 is quoted from Fig.3 in Carlin et al. (2012). It presents the $g-r$ vs $r-i$ diagram for different selections of stars in the region $(RA, Dec)=(130^\circ-136^\circ, 0^\circ-6^\circ)$, corresponding to a field center of $(l,b)\approx(225^\circ,28^\circ)$. Panel (a) shows all stars with $14^m<r<19.5^m$ in the field of view. Panel (b) shows the results of target selection with $\alpha=1$ and Panel (e) shows the results of target selection applied to the spheroid dark survey. Stars with $g-r<0.0^m$ in panel (b) are overemphasized by a slightly higher fraction than in panel (e).

\subsection{An estimate of the number of DAWDs in the pilot survey}
To ensure that we have identified all of the DAWDs with $S/N_g>10$, we estimate how many DAWDs are expected in the pilot survey. The target selection algorithm described in Section 2.2 provides a way to estimate the number of DAWDs.

To estimate the expected number of DAWDs we can find in the pilot survey, we discuss 4 cases separately,
\begin{itemize}
\item{dark plates in high Galactic latitudes, }
\item{dark plates in low Galactic latitudes, }
\item{bright plates in high Galactic latitudes,}
\item{bright plates in low Galactic latitudes.}
\end{itemize}

A $10^\circ\times10^\circ$ field is selected for each of the 4 cases from the input catalogs. Table 2 lists the equatorial and Galactic coordinates of the fields' centers, the number of stars in the fields and the corresponding input catalog, and the number of photometrically-selected DAWD candidates in the fields and that can be targeted in the pilot plates. Stellar photometry is taken from SDSS DR8. For the bright survey, the input catalog includes stars brighter than the magnitude limit of SDSS photometry. Because of the low brightness of white dwarfs, the missing bright stars have little effect on the estimation of DAWD numbers. The SEGUE white dwarf selection criteria (Yanny et al. 2009) are applied to get DAWD candidates. Here the DAWD candicates are obtained by the SDSS color cuts described in Yanny et al. (2009), which are different from the DAWD candidates discussed in Section 3 and 4. Magnitudes of stars in the high Galactic latitude fields are dereddened, while non-dereddened magnitudes are used for the low Galactic latitude fields where the extinction is strong; using the Schlegel extinction (Schlegel et al. 1998) at infinite distance, the magnitudes of targets in low Galactic latitudes will be significantly over-corrected. The input catalogs of the pilot survey are generated to be 600 stars deg$^{-2}$, i.e. 3 times of the fiber density in the focal plane, in order to make sure that all fibers are fed with targets. So not all the targets in the input catalog will be observed. The first 200 deg$^{-2}$ targets with the highest priorities are most likely to be observed. The number of DAWD candidates in the sub-sample of 20,000 stars (200 deg$^{-2}$ $\times$100 deg$^2$) with highest priorities are used to estimate the expected number of DAWDs in the pilot data release.

Figure 5 shows the $r$ magnitude distributions of the sample fields. The left panels are for all types of stars in the fields (solid), with the 200 deg$^{-2}$ target selection sub-samples shown in dashed lines. The right panels are for the DAWD candidates in the fields (solid) and for the target selection sub-samples (dashed). From top to bottom, they are high latitude dark field, low latitude dark field, high latitude bright field and low latitude bright field, respectively. The left panels show that stars brighter than $r=17.5^m$ are over-sampled in the dark fields as designed, while in the bright fields, the very bright and very faint stars are slightly overemphasized comparing to the majority of stars with $r\sim16^m$. The distributions of DAWD candidates in bright fields are not statistical because of the small samples.

Figure 6 shows the $g-r$ magnitude distributions of the sample fields corresponding to Figure 5. The stars with $g-r<1^m$ in the dark fields are strongly over-sampled as designed. In the bright fields rare stars are overemphasized while the most populated stars are relatively deemphasized. The clear excess of M-type stars at $g-r\sim1.5^m$ is caused by the wide spread of M-dwarf $r-i$ colors at nearly constant $g-r\sim1.5^m$, which dilutes the density of M stars in ($r$, $g-r$, $r-i$) space (Zhang et al. 2012). The $g-r$ colors of DAWD candidates locate at $-0.5^m<g-r<0.2^m$. The $g-r$ distribution of the targeted DAWDs in the dark fields is consistent with that of all DAWDs in the fields.

The 4 cases characterized in Table 2 are typical representatives of the respective survey areas, containing typical stellar populations and therefore WD statistics. Using the summary of plate numbers in Table 1 and considering the relative area of the 4 fields and the LAMOST survey footprint, we expect $\sim1,650$ DAWD candidates were targeted in the LAMOST pilot survey. According to Yanny et al. (2009), 62\% of the candidates are DAWDs down to $g=20.3^m$. Using the SEGUE color-cut white dwarf sample described in Yanny et al. (2009), we found that by cutting the white dwarf candidate sample at $g<19.5^m$, the success rate of DAWDs is $\sim60\%$. So we expect to find $\sim1,000$ DAWDs. A total of $\sim1,240,000$ stars were targeted on designed plates in the LAMOST pilot survey. Comparing to 640,000 spectra eventually obtained, 48\% spectra are lost because of instrumental malfunction or data reduction failure. Taking the 48\% spectra loss and the 53\% of spectra for analysis by constraining $S/N_g>10$ into account, $\sim290$ DAWDs can be expected from the pilot data release of LEGUE. Note that the scaling by 53\% is somewhat uncertain; since the $S/N$ selection might preferentially retain white dwarfs based on their bluer color than the general population, but also might preferentially throw out white dwarfs because they are fainter than the general population, we chose to simply use the fraction of stars retained by the $S/N$ selection.

\section{Algorithm for identifying DA white dwarfs}
\subsection{Balmer line profile fitting}

To search for DAWD spectra in a large spectral dataset, a prior selection of white dwarf candidates is helpful. Colors and proper motions have been used successfully in candidate selection of SDSS white dwarf catalogs (Harris et al. 2003, Kleinman et al. 2004, Eisenstein et al. 2006, Kleinman et al. 2013). However, the pilot survey input catalog is from various sources such as SDSS, 2MASS (the Galactic spheroid, see Yang et al. 2012 and Zhang et al. 2012), PPMXL (the disk, see Chen et al. 2012), and Xuyi photometric survey (the GAC, Yuan et al. 2013 in preparation). Applying different white dwarf classification criteria in different sky areas will lead to systematic classification inconsistency. Therefore, we are inclined to use only the WD spectral features to classify the WDs. Other information including colors, proper motions are used for reference only.

The best spectral features in the low resolution spectra that can be used to select WD from other stars are the profiles of the Balmer lines. Their shapes can be well fitted by a S\'ersic profile (S\'ersic 1968) usually used to fit the light profiles of an edge-on disk galaxy. In this context, Clewley et al. (2002) customized the S\'ersic function for line profiles,

\begin{equation}
y=n-a\times exp[-(\frac{|x-x_0|}{b})^c],\label{eq2}
\end{equation}

\noindent where $y$ is the flux of a normalized spectrum, $x$ the wavelength and $x_0$ the central wavelength of a Balmer line. The parameter $a$ indicates the depth of the line core, $b$ the line width and c the steepness of the line wing. The value of the parameter $n$, in principal, should be unity if the normalization of the continuum is done well. The parameter $b$ and $c$ are sensitive to surface gravity change in different targets, which were used by Clewly et al. (2002) and Xue et al. (2008), to successfully separate BHB stars from blue stragglers. The current work is to disentangle white dwarfs from other types of stars in similar parameter space. Since the best leverage is the surface gravity, we can therefore apply the same S\'ersic profile fitted parameters that are sensitive to gravity.

For high temperature targets such as WDs, the blue arm of LAMOST spectra is usually better than the red arm, therefore the line profile fitting is applied only to H$\beta$, H$\gamma$ and H$\delta$. The wavelengths of all the spectra are shifted to the rest frame. The central wavelengths of the lines are fixed at the vacuum wavelengths of H$\beta$, H$\gamma$ and H$\delta$. The entire spectrum is first normalized by the spline fitting. The spectrum is almost flat but the normalization is rough. So each of the Balmer line is then normalized locally by using pre-determined wavelength ranges on either side of the lines and fitting to a linear function. The wavelength ranges are chosen to be suitable for the broad Balmer lines in white dwarfs. The wavelength ranges for H$\beta$, H$\gamma$ and H$\delta$ local normalization are (4681 - 4741~\AA, 4981 - 5041~\AA), (4201 - 4241~\AA, 4441 - 4481~\AA) and (4031 - 4061~\AA, 4171 - 4201~\AA), respectively.

A nonlinear least square fitting is applied to each line to find the best fit parameters in Equation~(\ref{eq2}).
The root mean square (RMS) of the fitting and the coefficient of determination, $R^2$, are also calculated. The RMS value, which indicates the quality of the fitting, is affected by $S/N$ of the spectrum. $R^2$ also reflects the goodness of fitting in the sense of the linearity of the fitting. For a data set of values $y_i$, each of which has an associated modeled value $f_i$, $R^2$ is defined as

\begin{equation}
R^2 \equiv \frac{\sum (f_i-\overline{y})^2}{\sum (y_i-\overline{y})^2},\label{eqn}
\end{equation}

\noindent where $\overline{y}$ is the mean of $y_i$.
An $R^2$ close to unity indicates that the model data are very close to the real data and the model fits the data set well,  while an $R^2$ close to 0 means that the model does not fit the data. Figure 7 presents two examples of well fitted spectra. The upper panels show the Balmer lines (solid) and the best fit of S\'ersic profile (dashed) for an F star and the lower panels are for a DAWD.

The Balmer lines are prominent in spectra of stars within a surface temperature range around 10,000 K (A type stars). For stars with much higher (B type) or lower temperatures (G type and later), the Balmer lines are no longer dominant and even invisible. There are also QSOs in the input catalog. Since the Balmer line profile fitting may fail for these spectra, we first eliminated spectra without apparent Balmer lines, using constraints of RMS and $R^2$.
Figure 8 shows the correlation between $S/N_g$ and RMS of LAMOST pilot spectra with $S/N_g>10$. The colors of data points indicate the value of $R^2$. For spectra with the same $S/N_g$ value, the values of $R^2$ decrease as the values of RMS increase. We decided to use,
\begin{equation}
RMS<0.15, ~R^2>0.3,
\label{eq3}
\end{equation}

\noindent to keep well-fitted spectra of stars and to remove spectra without dominant Balmer lines. In order to avoid bad pixels that can affect the line profile by chance, Equation~(\ref{eq3}) should be satisfied by at least two of H$\beta$, H$\gamma$ and H$\delta$.

\subsection{DA white dwarf selection criteria}

The shapes of the Balmer lines of DAWDs depend on temperature and surface gravity. They are most prominent at DA4 subtype and weaken towards both ends of temperatures. They are broader for higher surface gravities. We use a SDSS field of spectroscopic objects as a test sample to get DAWD selection criteria. The sky coverage of the test sample is $120^\circ < RA < 250^\circ, 22^\circ < Dec < 65^\circ$. $(g-r)_0<0.4^m$ is applied to the sample to remove the large number of late type stars that have no dominant Balmer lines. The sample consists of 650 DAWDs, 121 subdwarfs, $\sim8000$ other types of stars and $\sim1500$ non-stellar objects from SDSS DR4 and the SDSS DR4 white dwarf catalog (Adelman-McCarthy et al. 2006, Eisenstein et al. 2006). A set of parameter constraints were applied to the test sample. The algorithm yielded very good performance when applied to the same data set, giving a 97.5\% efficiency for selecting DAWDs, and a relatively small contamination rate of 19.8\%.

To clarify the complicated selection criteria, we define two regions in the $a$-$b$ plane (Figure 9): Region 1 is defined by,
\begin{equation}0.1<a<0.8,~16<b<45.\label{eq4}\end{equation}

\noindent Region 2 is everywhere except for Region 1 on the plane, for all the three Balmer lines. Region 1 (defined for all the three Balmer lines) is where typical DAWDs should be located.  An object is considered to be in Region 1 whenever $a$, $b$ from any 2 of the lines are in the box defined by Equation~(\ref{eq4}), otherwise it is in Region 2.

Figure 9 demonstrates the two regions defined above. The data points come from the SDSS test sample. The red dots are main-sequence stars. The blue dots are DAWDs. And the green dots are subdwarfs. By applying  Equation~(\ref{eq3}), most of the non-stellar objects can be removed, with a few remaining shown as magenta crosses. In this plot, each object has 3 sets of parameters for H $\beta$, H $\gamma$ and H $\delta$ respectively. In Region 1, nearly all the targets can be identified as DAWDs. The constraint of $b$ selects typical high surface gravity DAWDs. The constraint of $a$ keeps the depth of the line core in an appropriate range.  In Region 2, white dwarfs are highly mixed with contaminants of subdwarfs and possibly A-type stars, therefore more constraints from CaII K line and $c$ are needed. $c$ indicates the steepness of the line wings, and CaII K lines are usually not present in DAWDs.

The equivalent width (EW) of the CaII K line is calculated. Similarly to the Balmer lines, the CaII K line is normalized locally. The wavelength range used for the linear fitting is (3920 - 3925~\AA, 3944 - 3948~\AA). The flux between 3927~\AA~ and 3942~\AA~ is used to calculate the EW of the CaII K line. Since the $S/N$ of the very blue end may be low and the wavelength range for normalizing the continuum is short, there may be negative values of the EW of CaII K line, which is not physical. When a negative EW is calculated, it is set to zero.

The criteria for WD selection in Region 2 are listed below, and must be satisfied by at least two of the Balmer lines:

\begin{eqnarray}
0.08<a<0.8,~0.6<c<1.8,\\
\label{eq5}
b<20,c<0.48~b-1.1,\\
a<1.0-0.35~c,\\
EW(CaII~K)<2,\\
~a<0.735-0.1~EW(CaII~K).
\label{eq6}
\end{eqnarray}

Figure 10 presents the selection criteria of Region 2. The symbols are the same as in Figure 9. The data points are only from Region 2 in Figure 9. The left panel is for targets on the $a$-$c$ plane, the middle panel on the $b$-$c$ plane. The right panel shows the EW of the CaII K line versus $a$.

\section{DA White Dwarfs from LAMOST pilot survey}

After applying the selection criteria to the LAMOST pilot data, we get 2964 DAWD candidates, of which 186 targets are from Region 1 and 2778 are from Region 2.

All candidate spectra are visually inspected and identified. Spectra are compared with the spectrophotometric atlas of Wesemael et al. (1993) and the spectra from the SDSS DR4 white dwarf catalog. Spectra with extremely broad H$\beta$, H$\gamma$ and H$\delta$ lines in Region 1 can be easily classified as DAWDs. For spectra with relatively narrow lines ($b<20$) or low $S/N$ ($S/N_g<15$), CaII H/K lines, CaII triplets and some helium lines are investigated to separate DAWDs from A-type stars and subdwarfs. For spectra with no line features in the wavelength range longer than H$\alpha$, intermediate line widths and weak CaII K line features that can not be recognized with confidence, it is difficult to identify them as DAWDs or subdwarfs, so we classify them as DAWD/subdwarfs. Some DAWDs are confirmed to have a non-degenerate companion, because of the bimodal profile of their continua or the different line widths between the blue and red part of the spectra. In the Appendix, the details of how to separate DAWDs from A-type stars and subdwarfs are presented.

Among 186 candidates in Region 1, 144 of them are confirmed as DAWDs, and 8 spectra as DAWD/subdwarfs. In Region 2, 86 of 2778 candidates are confirmed as DAWDs and 120 spectra as DAWD/subdwarfs. Most of the contaminants are subdwarfs and early type B/A stars. Because of the similar Balmer line profile, subdwarfs and early type stars pile up in Region 2. As a total, 230 DAWDs are identified, 20 of which are DA white dwarfs with non-degenerate companions. In addition, 128 candidates are identified as DA white dwarfs or subdwarfs stars. 59 of DAWDs are already in Villanova white dwarf catalog and SDSS WD catalog. One of them is a DAWD with a M-type secondary star classified as ``DA2:+M".

Table 3 lists DAWDs and DAWD/subdwarf targets. LAMOST ID is made up of RA and Dec of J2000. $u$, $g$, $r$, $i$ magnitudes come from SDSS photometry or Xuyi survey. J, H, K magnitudes come from 2MASS photometry. B, R and I magnitudes come from UCAC3 photometry. The V magnitude is from Hipparcos catalog (Perryman et al. 1997). The note ``v" or ``s" is presented if the target has been identified in Villanova catalog (``v") or SDSS white dwarf catalog (``s"). According to Sion et al. (1983), the classification consists of two parts, the ``DA" type and the subtype index which is a temperature index corresponding to $50,400/T$. This temperature index is determined by visually comparing the target spectrum with spectra of different temperature indices in the atlas of Wesemael et al. (1993) and SDSS DR4 white dwarf catalog. A DAWD with a companion is denoted as ``DA+companion", if the spectral type of the companion is identifiable, i.e. a M star, the classification is ``DA+M". A colon means the identification is uncertain. The DAWD/subdwarf targets are classified as ``DA:/SD".

An atlas of DAWD spectra is generated, and the Balmer line profile fittings of all DAWD spectra are plotted. Figure 11 and Figure 12 present examples of the atlas and the Balmer lines. As discussed above, the temperature indices are assigned by visual inspection. The full atlas is available in the online version.

Figure 13 compares the LAMOST and SDSS spectra for the same DAWDs. There are 28 DAWDs that have also been observed by SDSS. Here we select three targets to show different image quality of LAMOST spectra. From top to bottom, the $g$ magnitudes of the three DAWDs are $g = 15.31$, $16.03$ and $16.22$, with LAMOST g-band $S/N$ ratios of 45.7, 19.5, and 13.7, respectively. The black lines are LAMOST spectra and the red lines are SDSS spectra. The stellar parameters, magnitudes, classifications, and IDs of the DAWDs are also presented in the plots. The top panel shows that the spectra of LAMOST and SDSS are almost the same when LAMOST achieves adequate $S/N$. The middle panel shows that the continuum of the LAMOST spectrum is consistent with the SDSS spectrum, but with lower $S/N_g$ and worse sky subtraction in the red part. The continuum of the LAMOST spectrum in the bottom panel is clearly different from that of the SDSS spectrum. This is likely because the low $S/N$ at the blue end affected the flux calibration.

\section{Discussion and conclusion}

The positions of DAWDs are presented in Figure 1 as red dots (bright survey) and green dots (dark survey). 247 DAWDs and DAWD/subdwarfs are from bright survey and 112 from dark survey (one of them was targeted by both the bright and dark survey). 80\% of the bright targets are located at the low Galactic latitude disk and GAC region. The faint targets are distributed sparsely in the dark plates.

The number of DAWDs we found ($\sim240-350$) is consistent with the expected DAWD numbers ($\sim290$). Figure 14 shows the $g-r$ distribution (left) and $r$ distribution (right) of all released pilot spectra and DAWDs we found. The gray solid lines represent all spectra in pilot survey. The gray dashed lines represent all DAWD and DAWD/subdwarf spectra. The black solid line represents single DAWDs and the black dashed line represents DAWDs with non-degenerate companions. The $g$ and $r$ magnitudes are from cross-identification within a 5" radius of PanSTARRS and UCAC4. A few of the DAWDs with $g-r>0.5^m$ might be mistakes of cross-identification or they are heavily reddened stars. The reason this can not be clarified currently is because of the variety of input catalog sources. If the red colors are due to extinction, finding white dwarfs by their spectra, then using multi-band synthetic magnitudes to derive Galactic extinction would be valuable. This has been done in Holberg et al. (2008) for the Eisenstein SDSS DR4 white dwarf sample, and can be applied to LAMOST white dwarfs in the future. The color distribution of single DAWDs is similar to those shown in Figure 6. The $r$ magnitude distribution shows two peaks at $r\sim14^m$ and $r\sim16.5^m$ similar to the $r$ distribution of spectra with $S/N_g>10$ of the bright and dark survey respectively. So the magnitude limit of DAWDs is decided by the limiting magnitude of the telescope. Figure 15 shows the $u-g$ vs $g-r$ 2-color diagram (left) and $g-r$ vs $r$ color-magnitude diagram (right) of pilot spectra and DAWDs that have SDSS $ugr$ photometry. The black/gray areas represent all spectra. The red triangles are ``DA:/SD" targets. The green dots are single DAWDs. The magenta pentacles are DAWDs with companions. Most of the single DAWDs and DA binaries locate clearly in the expected region in $u-g$ vs $g-r$ diagram.

Some special DAWDs are found in LAMOST pilot data. There are 20 DAWDs with a companion. Some of spectra have a bimodal continuum, indicating a hot DAWD in a binary system with a late type star. The continua of some other spectra are not significantly bimodal, but the line widths of shorter wavelength (H$\gamma$, H$\beta$) and longer wavelength lines (H$\alpha$, sometimes CaII triplets) are different. It means the white dwarfs may have non-degenerate companions with slightly redder color. Figure 16 shows examples of the two features above.

There are several interesting spectra that have emission lines in the cores of absorption lines of H$\alpha$ and even H$\beta$. Most of these spectra are targets classified as ``DA:/SD". They may not be DAWDs. But these stars may undergo an interesting process. Figure 17 presents some of these spectra with emission lines.

The arrangement of bright and dark plates in the pilot survey is mainly determined by the weather conditions and Moon phase at the LAMOST site. In 5 years of LAMOST regular survey, LEGUE will observe more than 7.5 million stars in the Galaxy $-$ 6 times larger than observed in the pilot survey. If the magnitude limits can be extended down to 19.5 after further tuning of the instruments, $\sim1,000$ DA white dwarf spectra can be obtained in a year. Spectra of at least 7,000 DAWDs will be obtained with more than 60\% of them are in the disk and Galactic anticenter region. Note that $\sim13,000$ DAWD spectra were obtained from SDSS DR7, twice the number expected from LAMOST. This is due to the different target selection method. SEGUE selected all white dwarf candidates by the color selection criteria in Yanny et al. (2009), while LAMOST overselects rare targets by a well-defined selection algorithm. In sky areas with much higher stellar number density, $\sim30\%$ of DAWD candidates are targeted (see the results for high Galactic latitude dark case and low Galactic latitude dark case in Table 2). However, with the known target selection function, the color and magnitude distribution of all DAWDs can be estimated. The distribution of the overall sample of DAWDs in a certain sky area will be obtained. However, when the DAWD candidate sample becomes much larger, the visual inspection and identification will be inefficient. An automatic numerical process, like the SDSS white dwarf template matching method, is needed. Such a technique will allow us to obtain the stellar parameters of the white dwarfs.

Eventually, we will develop an algorithm to carry out matching of template spectra based on model atmosphere grids to derive atmospheric parameters (T$_{eff}$, log g, mass, age, etc.) for the DAWDs we identified by our technique. When combined with kinematic information such as the radial velocity and proper motion, the catalog of DAWDs in LEGUE survey will be a valuable mine to study the formation, evolution and kinematics of the Galactic disk and disentangle the mixture of the thin disk, the thick disk and the spheroid populations.

\appendix
\section{Spectral features used to identify a DAWD}

The typical feature of DAWDs is broad Balmer lines. For Balmer lines of higher level than H$\gamma$, the lines become weaker due to the blanketing effect. However, the line width of DAWDs of lower surface gravity are similar to A-type stars and subdwarfs. And the line width of DAWDs with lower temperatures are narrower, as the hydrogen lines narrow with decreasing temperatures because of the Stark effect (Gray \& Corbally 2009). In this case, the CaII H/K and CaII triplets are investigated. Spectra with strong CaII K are classified as A stars and removed from the candidate list. In the early A-type stars, CaII triplets are weaker than the Paschen lines, but in late A-type stars the CaII triplets become dominant. It can also be a feature to recognize A-type stars. The spectra of subdwarfs are with no clear metallic lines, relatively narrow Balmer lines ($b<15$) and narrow hellium lines for some of them.

Here we present some spectra examples to show how DAWDs are separated from A-type stars and subdwarfs in Figure 18.
The first plot is the spectrum of an A-type star. It shows CaII K line at the blue end and CaII triplets in the red end. The H$\epsilon$ seems much stronger than other Balmer lines because the CaII H line is mixed with it. The second plot is the spectrum of a subdwarf. It shows HeI $\lambda\lambda4026$, 4387, 4471~\AA~ and the CaII K line. The third plot is a ``DA3:" white dwarf. The sky subtraction in the red end is problematic but the broad Balmer lines and the weakening line depth of higher level lines indicate it is a DAWD. The spectrum in the fourth plot exhibits only weak Balmer lines of H$\alpha$, H$\beta$, H$\gamma$ and H$\delta$. The fact that the higher level Balmer lines are weaker than the lower level lines convinces us it is a DAWD with low temperature, a ``DA8:" white dwarf.

\acknowledgments{We thank the referee, Dr. Jay Holberg, for helpful comments and suggestions on the manuscript. Guoshoujing Telescope (the Large Sky Area Multi-Object Fiber Spectroscopic Telescope LAMOST) is a National Major Scientific Project built by the Chinese Academy of Sciences. Funding for the project has been provided by the National Development and Reform Commission. LAMOST is operated and managed by the National Astronomical Observatories, Chinese Academy of Sciences.  This work is supported in part by NSFC through grants 10973015. CL is supported by NSFC grant U1231119. YX is supported by NSFC grant Y111221001. JLC and HJN are supported by NSF grant AST 09-37523. }

\clearpage
\begin{table}
\caption{LAMOST pilot survey plate statistics}
\begin{tabular}{ccccc}
\hline
\multicolumn{2}{c}{region} & No. of dark plates & No. of bright plates & No. of all plates\\
\hline
\multirow{2}{*}{spheroid} & $|b|>30^\circ$ & 57 & 112 & 169\\
& $|b|<30^\circ$ & 5 & 11 & 16\\
\hline
\multirow{2}{*}{GAC} & $|b|>30^\circ$ & 4 & 6 & 10\\
& $|b|<30^\circ$ & 50 & 103 & 153\\
\hline
\multirow{2}{*}{disk} & $|b|>30^\circ$ & 0 & 0 & 0\\
& $|b|<30^\circ$ & 2 & 31 & 33\\
\hline
\multirow{2}{*}{all regions} & $|b|>30^\circ$ & 61 & 118 & 179\\
& $|b|<30^\circ$ & 57 & 145 & 201\\
\hline
total & all Galactic b & 118 & 263 & 381\\
\hline
\end{tabular}
\end{table}

\begin{deluxetable}{ccccc}
\tabletypesize{\scriptsize}
\tablewidth{0pt}
\tablecolumns{5}
\tablecaption{Estimate of DA white dwarf candidates in the pilot survey}
\tablehead{
\colhead{Case} & \colhead{high Gb, dark} & \colhead{low Gb, dark} & \colhead{high Gb, bright} & \colhead{low Gb, bright} }
\startdata
(RA, Dec) & (195, 57) & (127, 4) & (160, 29) & (337, 29) \\
(l, b) & (120.6,60.1) & (220.7,23.3) & (201.4,61.0) & (88.7,-24.2) \\
stars(field) & 108396 & 421311 & 36740 & 156395 \\
stars(input) & 57296 & 60000 & 40454 & 60000 \\
stars(input) with $ugir$  & 57296 & 60000 & 26678 & 50936 \\
DAWD candidates(field) & 176 & 217 & 10 & 32 \\
DAWD candidates(input) & 170 & 138 & 4 & 7 \\
DAWD candidates (200 deg$^{-2}$) & 67 & 61 & 4 & 2 \\
\enddata
\end{deluxetable}

\clearpage
\begin{deluxetable}{cccccccccccccccl}
\tabletypesize{\scriptsize}
\tablewidth{0pt}
\rotate
\tablecolumns{15}
\tablecaption{Identified DAWDs and DAWD/subdwarf targets in LAMOST pilot survey}\label{Tab:DAWD}
\tablehead{
\colhead{No.} & \colhead{LAMOSTID} & \colhead{$u$} & \colhead{$g$} & \colhead{$r$} & \colhead{$i$} & \colhead{$J$} & \colhead{$H$} & \colhead{$K$} & \colhead{$B$} & \colhead{$R$} & \colhead{$I$} & \colhead{$V$} & \colhead{RV} & \colhead{class} & \colhead{note}}
\startdata
1 & J004036.79+413138.7 & \nodata & 15.9 & 16.21 & 16.4 & \nodata & \nodata & \nodata & \nodata & \nodata & \nodata & \nodata & 111.45 & DA3 & \nodata \\
2 & J004139.06+420713.5 & \nodata & 16.28 & 16.84 & 17.11 & \nodata & \nodata & \nodata & \nodata & \nodata & \nodata & \nodata & 20.55 & DA: & \nodata\\
3 & J004425.49+022244.7 & 17.59 & 16.56 & 16.57 & 16.61 & \nodata & \nodata & \nodata & \nodata & \nodata & \nodata & \nodata & -172.26 & DA5: & \nodata\\
23 & J015259.14+010016.8 & 17.11 & 16.41 & 16.58 & 16.76 & \nodata & \nodata & \nodata & \nodata & \nodata & \nodata & \nodata & 25.91 & DA2: & v,s\\
24 & J020606.40+585528.9 & \nodata & \nodata & \nodata & \nodata & \nodata & \nodata & \nodata & \nodata & \nodata & \nodata & 13.59 & -48.84 & DA5: & \nodata\\
25 & J021816.56+603619.0 & \nodata & \nodata & \nodata & \nodata & \nodata & \nodata & \nodata & \nodata & \nodata & \nodata & 13.67 & -43.94 & DA:/SD & \nodata\\
26 & J022454.05+565936.9 & \nodata & \nodata & \nodata & \nodata & \nodata & \nodata & \nodata & \nodata & \nodata & \nodata & 14.4 & -55.43 & DA:/SD\tablenotemark{e} & \nodata\\
71 & J035710.48+284941.6 & \nodata & 17.18 & 17.59 & 17.73 & \nodata & \nodata & \nodata & \nodata & \nodata & \nodata & \nodata & -17.73 & DA:/SD & \nodata\\
72 & J035729.86+501025.4 & \nodata & \nodata & \nodata & \nodata & \nodata & \nodata & \nodata & \nodata & \nodata & \nodata & 13.96 & 54.68 & DA: & \nodata\\
73 & J035838.14+461656.8 & \nodata & \nodata & \nodata & \nodata & 11.19 & 10.39 & 10.11 & \nodata & \nodata & \nodata & \nodata & -23.54 & DA1-2: & \nodata\\
74 & J035845.18+484553.6 & \nodata & \nodata & \nodata & \nodata & \nodata & \nodata & \nodata & 15.8 & 15.12 & 14.86 & \nodata & -28.5 & DA:/SD\tablenotemark{e} & \nodata\\
75 & J035908.87+451201.2 & \nodata & \nodata & \nodata & \nodata & 12.18 & 11.49 & 11.27 & \nodata & \nodata & \nodata & \nodata & -50.25 & DA1-2:+companion & \nodata\\
351 & J192207.07+370546.0 & \nodata & \nodata & 13.7 & \nodata & \nodata & \nodata & \nodata & \nodata & \nodata & \nodata & \nodata & -6.94 & DA: & \nodata\\
352 & J192240.10+370318.4 & \nodata & \nodata & 13.6 & \nodata & \nodata & \nodata & \nodata & \nodata & \nodata & \nodata & \nodata & -83.06 & DA: & \nodata\\
353 & J192337.92+395957.5 & \nodata & \nodata & \nodata & \nodata & \nodata & \nodata & \nodata & \nodata & \nodata & \nodata & 12.24\tablenotemark{k} & -31.17 & DA1-2 & \nodata\\
354 & J193205.44+394456.6 & \nodata & \nodata & \nodata & \nodata & \nodata & \nodata & \nodata & \nodata & \nodata & \nodata & 13.82\tablenotemark{k} & -75.85 & DA2: & \nodata\\
\enddata
\tablenotetext{e}{The spectrum shows emission.}
\tablenotetext{k}{The magnitude is $K_p$ magnitude from Kepler Input Catalog (Brown et al. 2011). These targets are in two plates that were put forward after the survey begin.}
\tablecomments{ ~LAMOSTID indicates RA and Dec of J2000. $u$, $g$, $r$, $i$ magnitudes come from SDSS photometry or Xuyi survey. $J$, $H$, $K$ magnitudes come from 2MASS photometry. $B$, $R$ and $I$ magnitudes come from UCAC3 photometry. $V$ magnitude is from Hipparcos catalog. The note ``v" means a DAWD is in Villanova white dwarf catelog and ``s" means in SDSS DR7 white dwarf catalog. Table 3 is published in its entirety in the electronic edition of the Astronomical Journal. A portion is shown here for guidance regarding to its form and contents.}
\end{deluxetable}

\begin{figure}
\plotone{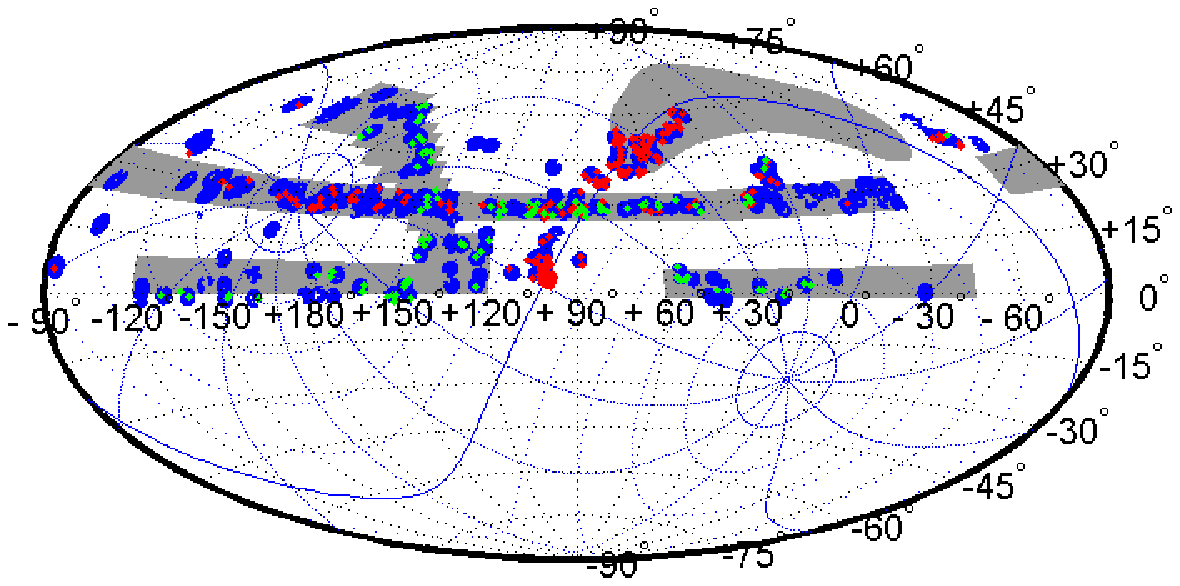}
\figcaption{The spatial distribution of targets of LAMOST pilot survey. The blue dots are targets observed. The gray areas are the designed input catalog. Note that several plates are not within the sky area of the designed footprint, these plates were put forward after the survey began. The red dots are DAWDs in bright plates and the green dots DAWDs in dark plates.}
\end{figure}

\begin{figure}
\epsscale{1}
\plotone{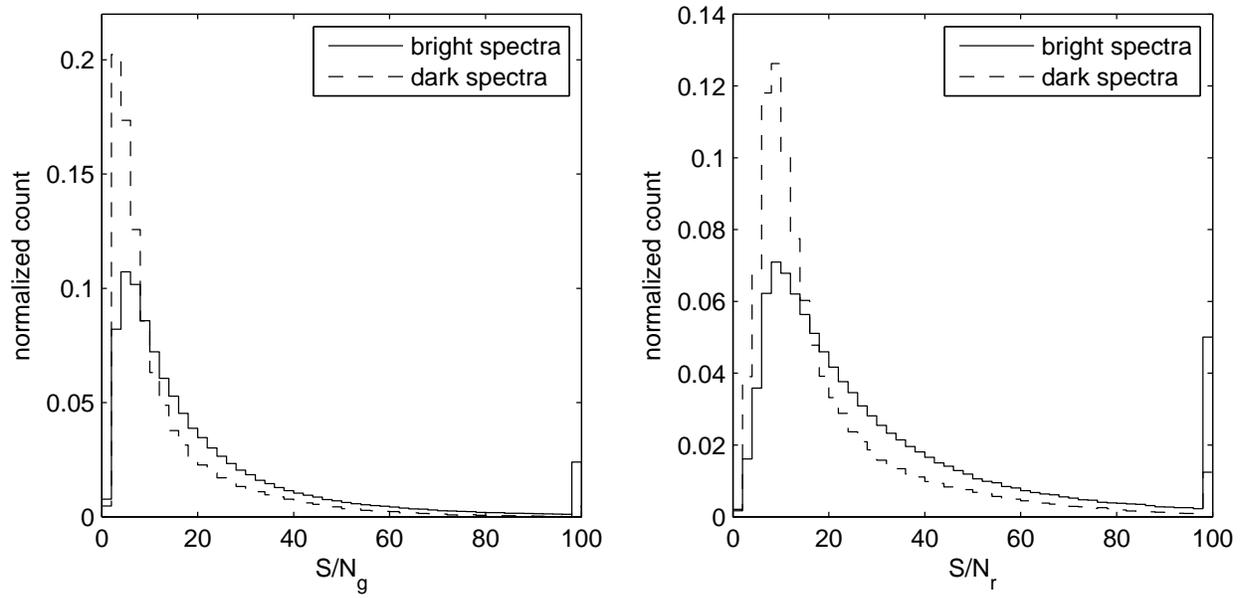}
\figcaption{The distributions of $S/N$ in $g$ and $r$ band of spectra in bright (solid) and dark plates (dashed).}
\end{figure}

\begin{figure}
\epsscale{0.8}
\plotone{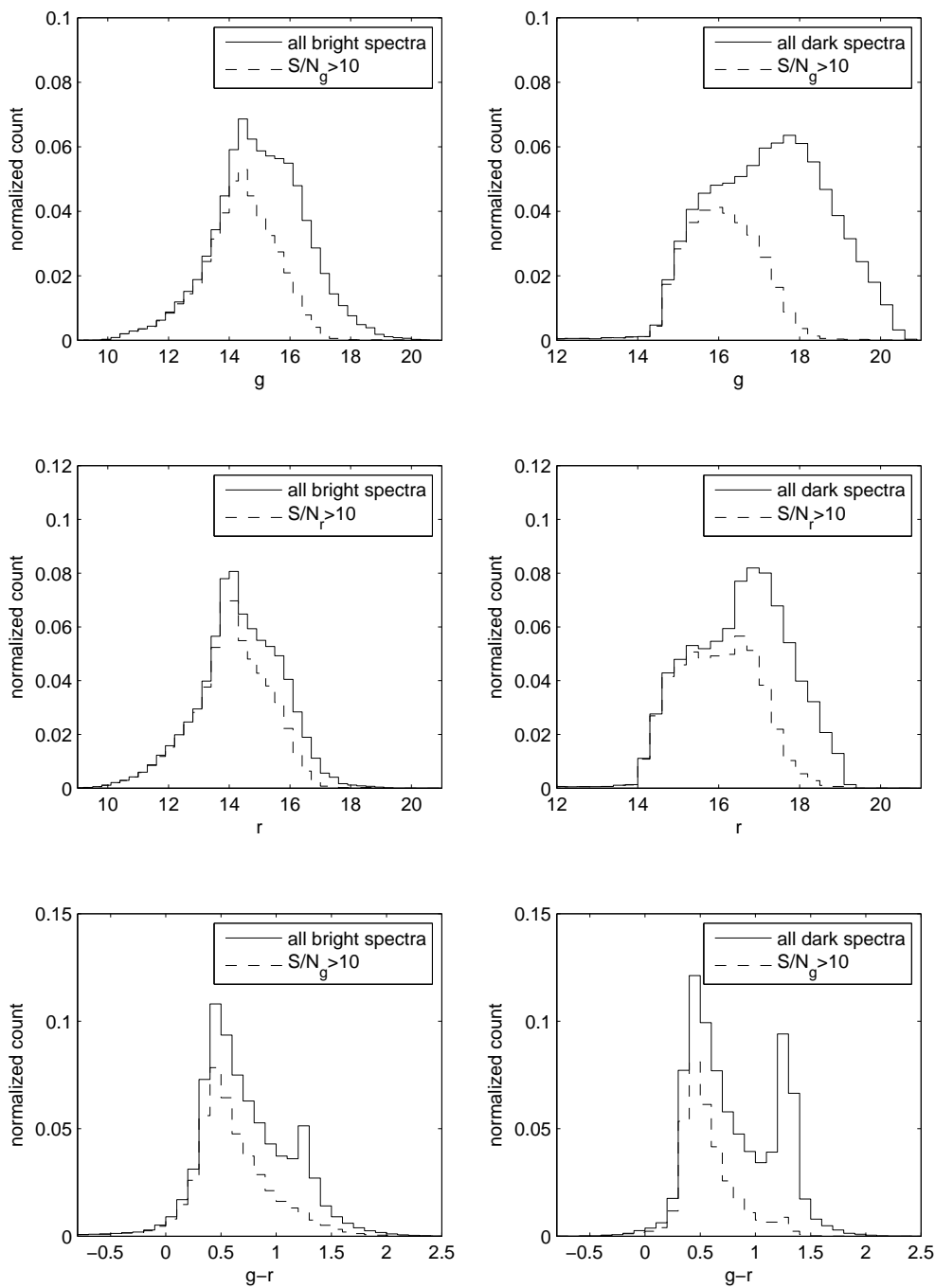}
\figcaption{The distributions of $g$ and $r$ magnitude and $g-r$ color of spectra in bright (left panels) and dark plates (right panels). The distributions of the sub-samples with $S/N_g>10$ or $S/N_r>10$ are also plotted (dashed). }
\end{figure}

\begin{figure}
\plotone{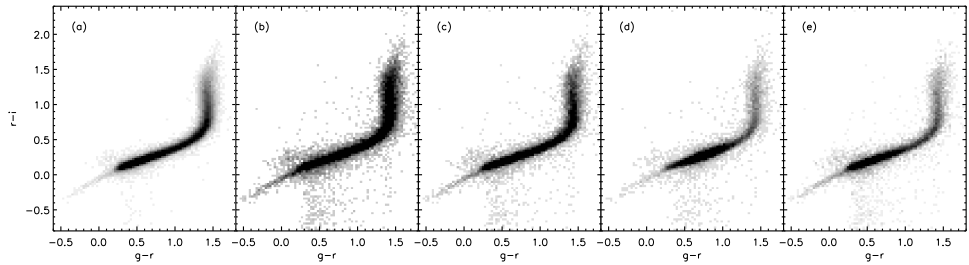}
\figcaption{Quoted from Fig.3 in Carlin et al. (2012). It presents the $g-r$ vs $r-i$ diagram for different selections of stars in the region $(RA, Dec)=(130^\circ-136^\circ, 0^\circ-6^\circ)$, corresponding to a field center of $(l,b)\approx(225^\circ,28^\circ)$. Panel (a) shows all stars with $14^m<r<19.5^m$ in the field of view. Panel (b) shows the results of target selection with $\alpha=1$ and Panel (e) shows the results of target selection applied to the spheroid dark survey. }
\end{figure}

\begin{figure}
\epsscale{0.8}
\plotone{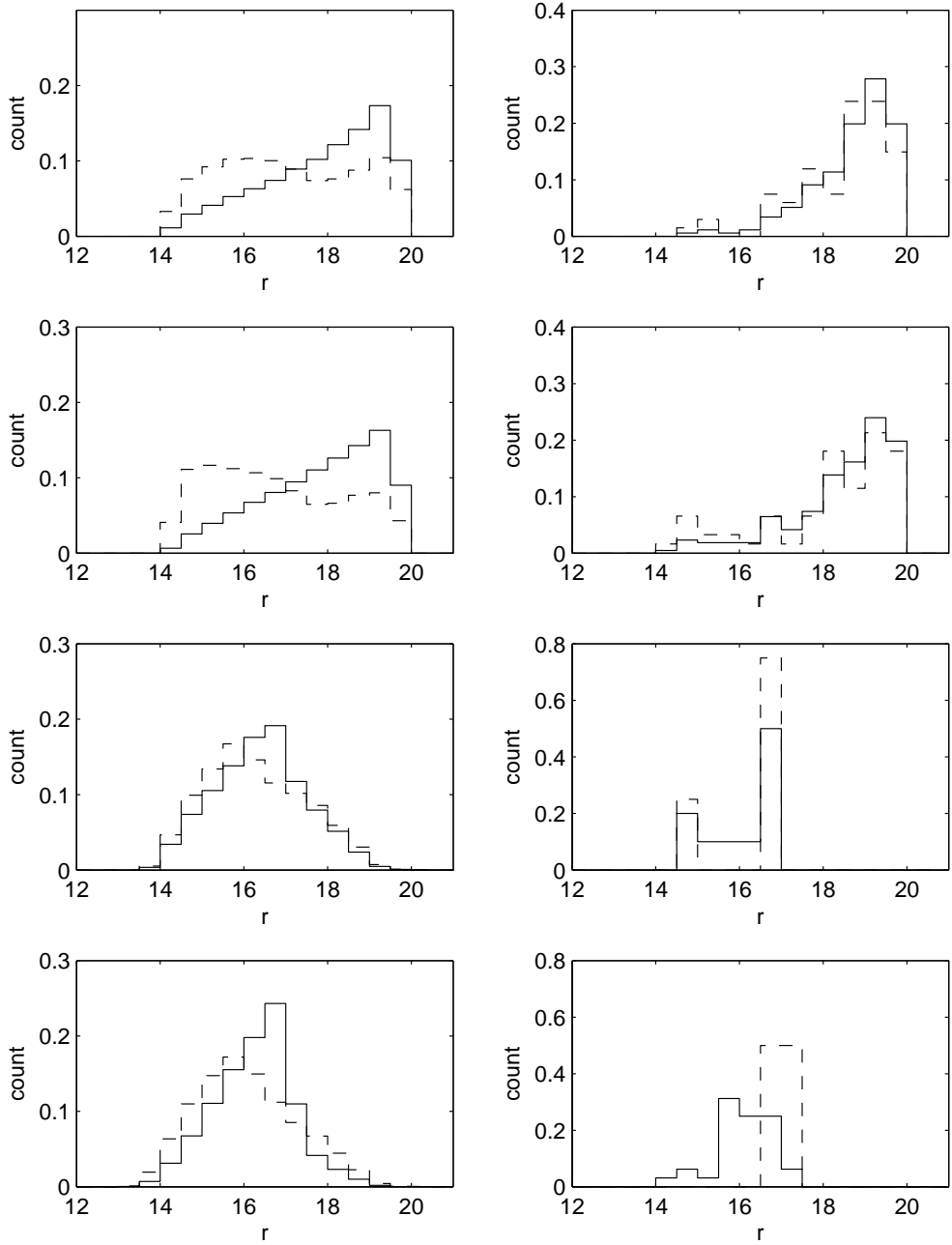}
\figcaption{The $r$ magnitude distributions of the field samples. The left panels are for all types of stars in the fields (solid), with the 200 deg$^{-2}$ target selection sub-samples shown in dashed lines. The right panels are for the DAWD candidates in the fields (solid) and for the target selection sub-samples (dashed). From top to bottom, they are high latitude dark field, low latitude dark field, high latitude bright field and low latitude bright field, respectively.}
\end{figure}

\begin{figure}
\plotone{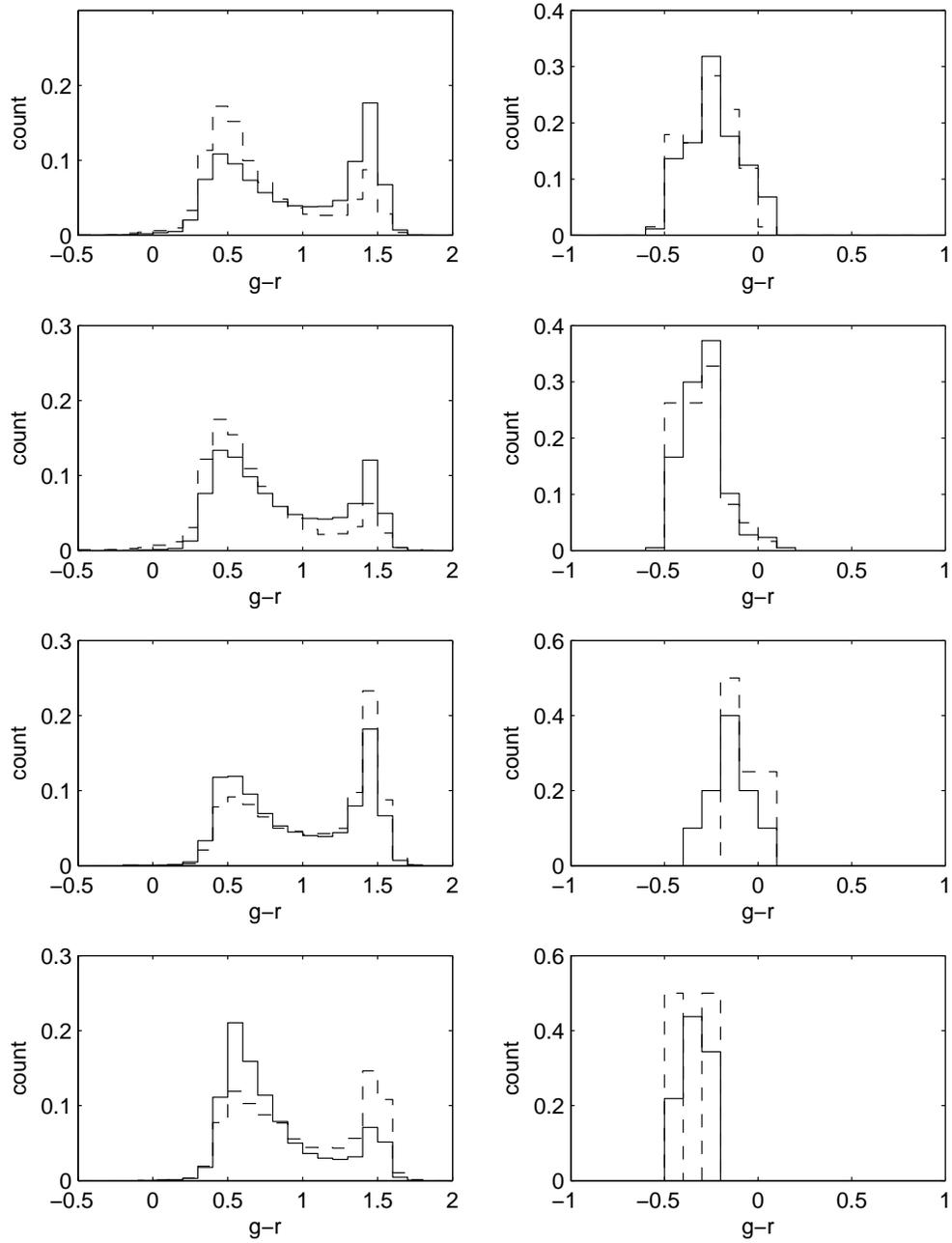}
\figcaption{The color $g-r$ distributions of the field samples corresponding to Figure 5. }
\end{figure}

\begin{figure}
\plotone{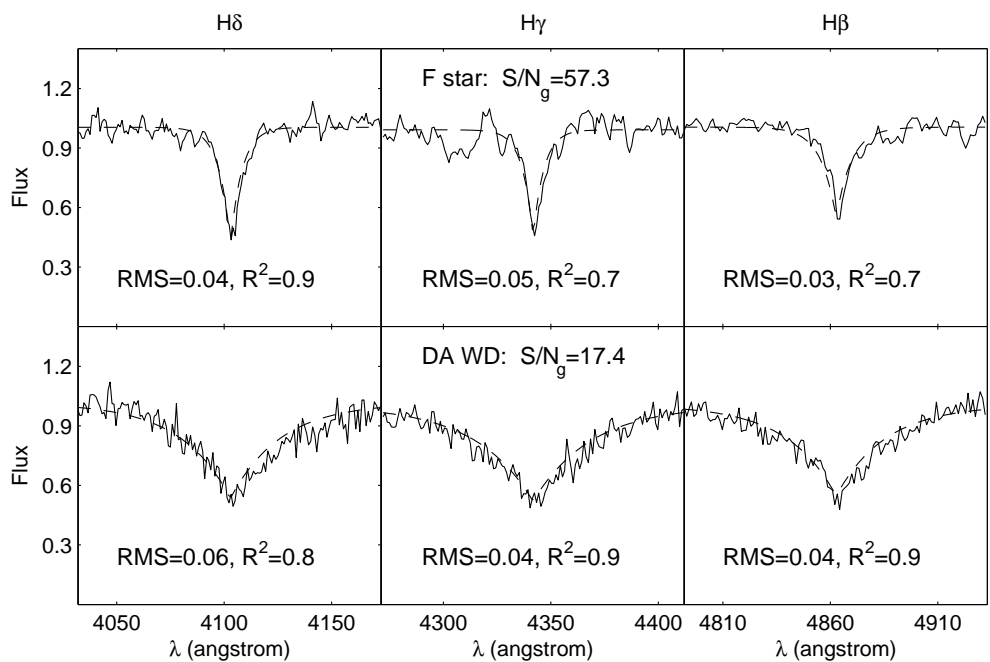}
\figcaption{Examples of Balmer line profile fitting. The upper panels show the Balmer lines (solid) and the best fit of S\'ersic profile (dashed) for a F star and the lower panels are for a DAWD.}
\end{figure}

\begin{figure}
\plotone{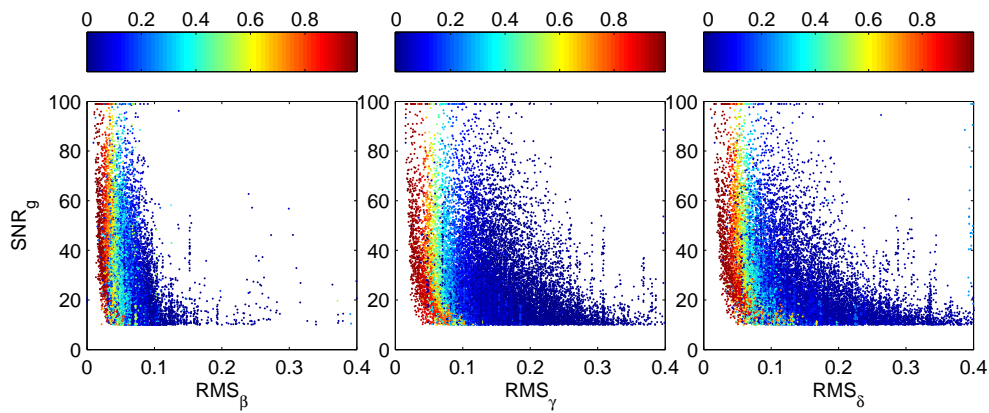}
\figcaption{The correlation between $S/N_g$ and RMS. The colors of data points indicate the value of $R^2$. For spectra with the same $S/N_g$ value, the values of $R^2$ decrease as the values of RMS increase. }
\end{figure}

\begin{figure}
\epsscale{1}
\plotone{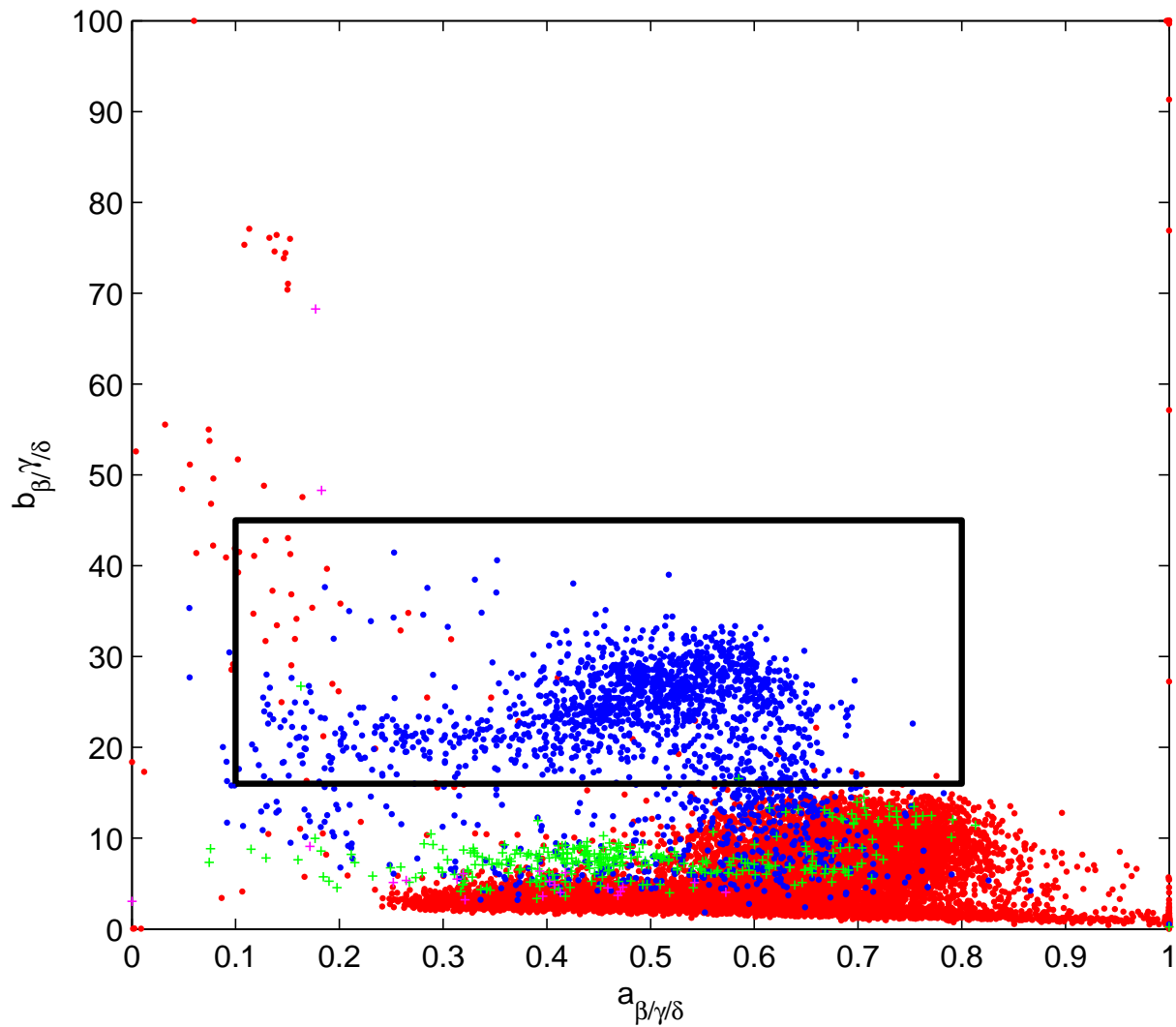}
\figcaption{All data points are parameter $a$ vs. $b$ for fitting of H$\beta$, H$\gamma$ and H$\delta$ from the SDSS testing sample. The red dots are main-sequence stars. The blue dots are DAWDs. The green dots are subdwarfs. And the magenta cross are non-stellar objects, most of which have been removed by the constraints of RMS and $R^2$. Each object has 3 sets of parameters of H$\beta$, H$\gamma$ and H$\delta$. The black box is the boundary of Region 1 defined by Equation~(\ref{eq4}).}
\end{figure}

\begin{figure}
\epsscale{}
\plotone{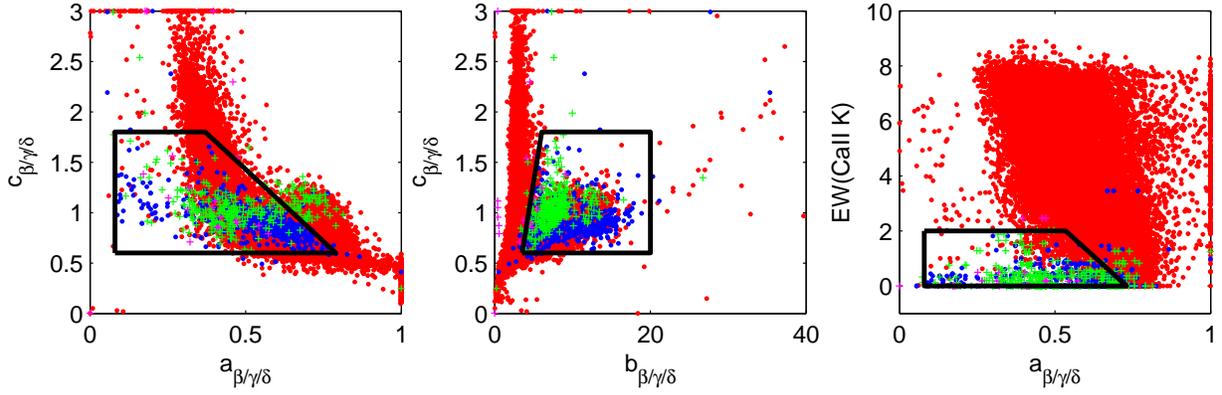}
\figcaption{Symbols are the same as in Figure 9. The data points are only from Region 2. The black boxes are projections of DAWD selection criteria in Region 2 defined by Equation~(\ref{eq5})-(\ref{eq6}) in the $a$-$c$, $b$-$c$ and $a$-EW(CaII K) planes, respectively.}
\end{figure}

\begin{figure}
\plotone{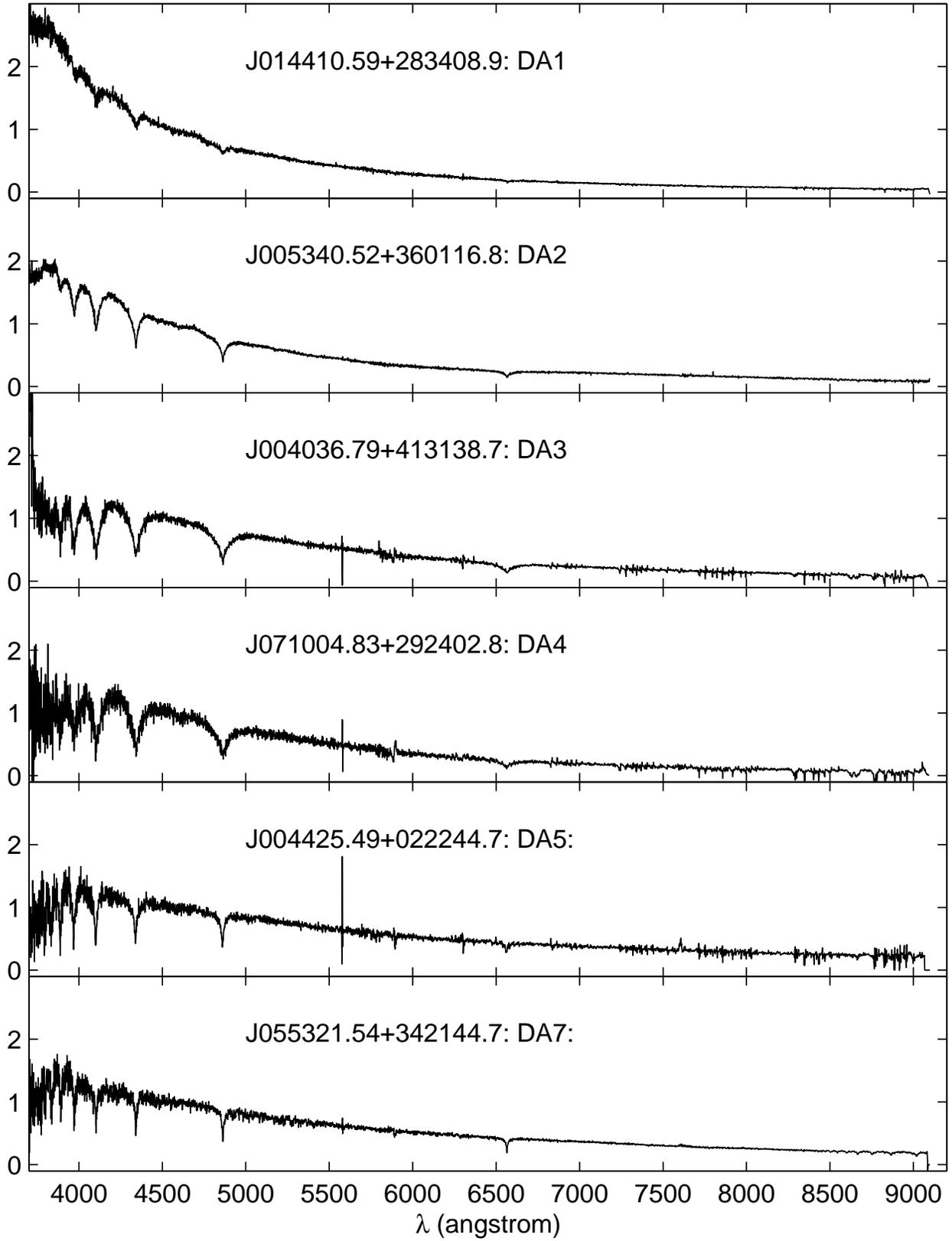}
\figcaption{Examples of DAWDs with subtype 1-7. The spectra are scaled by the mean flux between 4500 and 4600~\AA.}
\end{figure}

\begin{figure}
\plotone{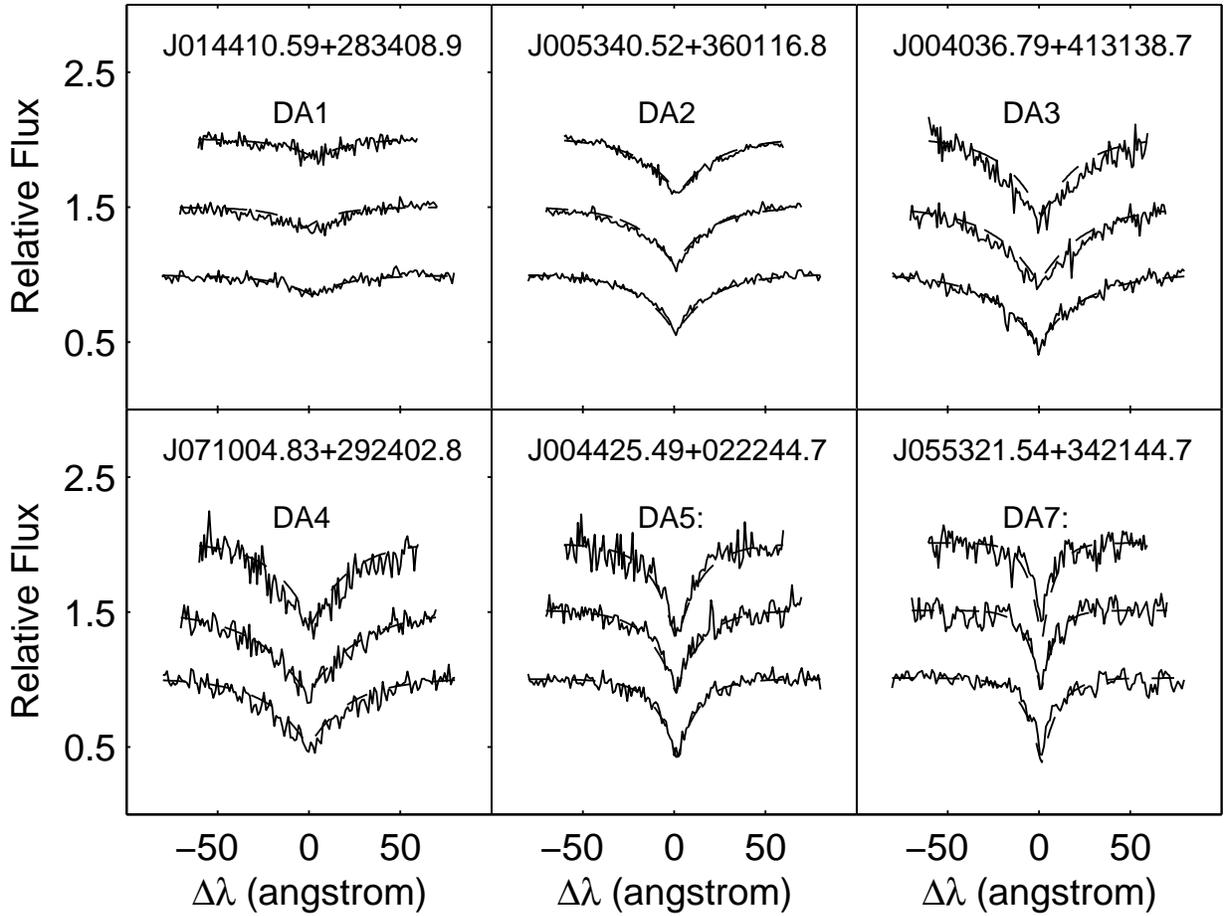}
\figcaption{Examples of Balmer line profile fitting of the DAWD spectra presented in Figure 11.}
\end{figure}

\begin{figure}
\plotone{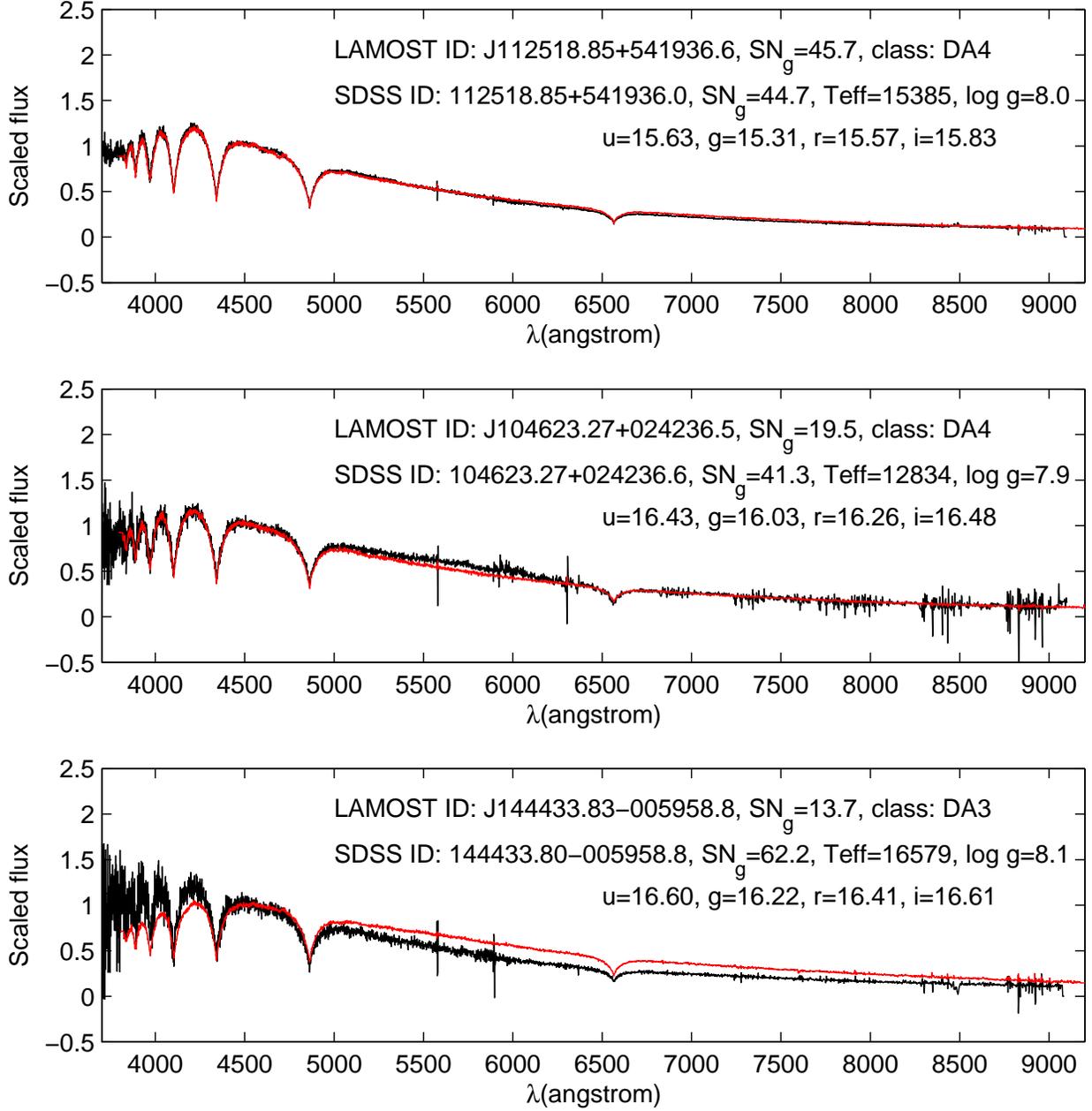}
\figcaption{ Examples of the LAMOST and SDSS spectra observed for the same DAWDs.  The three stars were selected to show the range of quality (i.e., $S/N$) among the LAMOST pilot survey DAWDs. The stellar parameters, magnitudes, spectroscopic $S/N$, classifications, and IDs of each star are presented in the plots. The black lines are LAMOST spectra and the red lines are SDSS spectra. The spectra are scaled by the mean flux between 4500 and 4600~\AA. The high-$S/N$ LAMOST spectrum in the top panel is nearly identical to the SDSS spectrum, with lower $S/N$ stars in the lower two panels showing clear differences in their flux calibrations.}
\end{figure}

\begin{figure}
\plotone{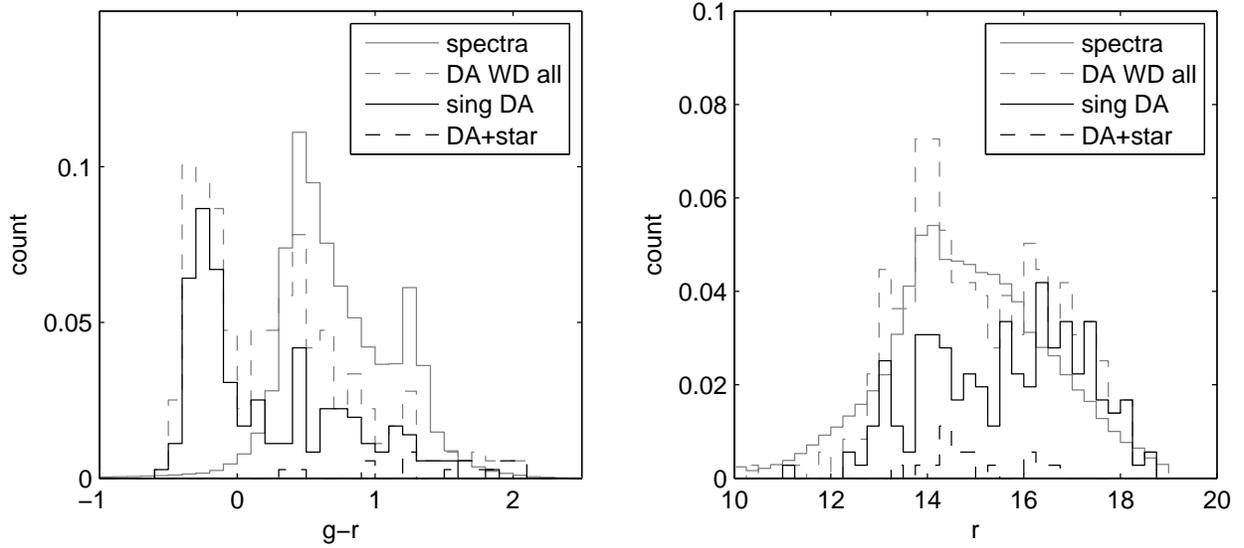}
\figcaption{The $g-r$ color (left) and $r$ magnitude (right) distributions of DAWDs found comparing to all spectra. The gray solid lines represent all spectra in pilot survey. The gray dashed lines represent all DAWD and DAWD/subdwarf spectra. The black solid line represent single DAWDs and the black dashed line represent DAWDs with non-degenerate companions.}
\end{figure}

\begin{figure}
\plotone{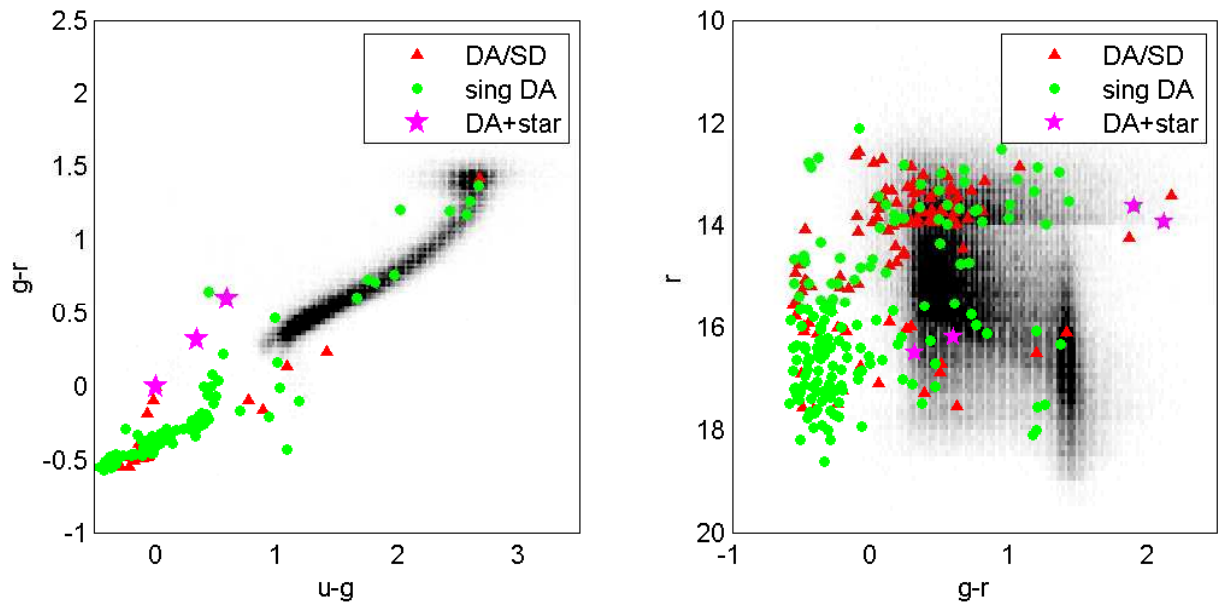}
\figcaption{The $u-g$ vs $g-r$ 2-color diagram (left) and the color-magnitude diagram of $g-r$ vs $r$ (right) of DAWDs comparing to all spectra. The black/gray areas represent all spectra. The red triangles are ``DA:/SD" targets. The green dots are single DAWDs. The magenta pentacles are DAWDs with companions.}
\end{figure}

\begin{figure}
\plotone{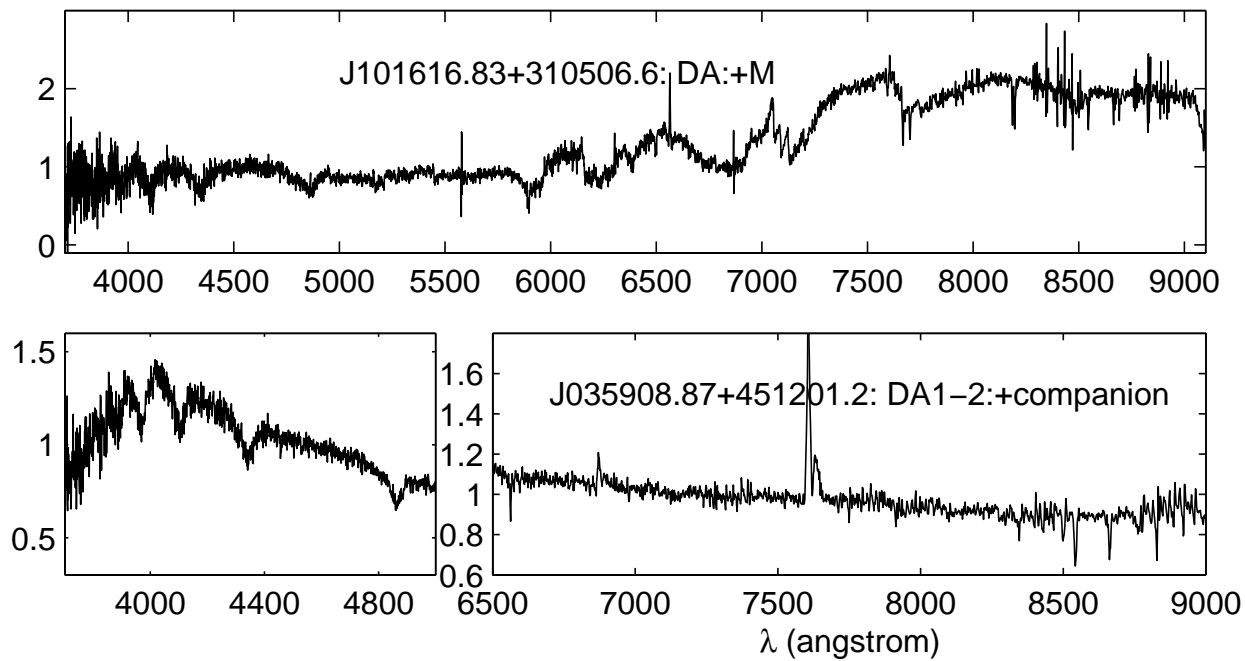}
\figcaption{Examples of DAWDs with non-degenerate companions. The upper spectrum and the blue part of the lower spectrum are scaled by the mean flux between 4500 and 4600~\AA. The red part of the lower spectrum is scaled by the mean flux between 7000 and 7500~\AA. The upper spectrum shows significant bimodal continuum features as a DAWD with a M star companion. The lower one shows a uniform continuum profile but the Balmer lines in the blue part are wide enough to be a DAWD while H$\alpha$ line in the red part is too narrow.}
\end{figure}

\begin{figure}
\plotone{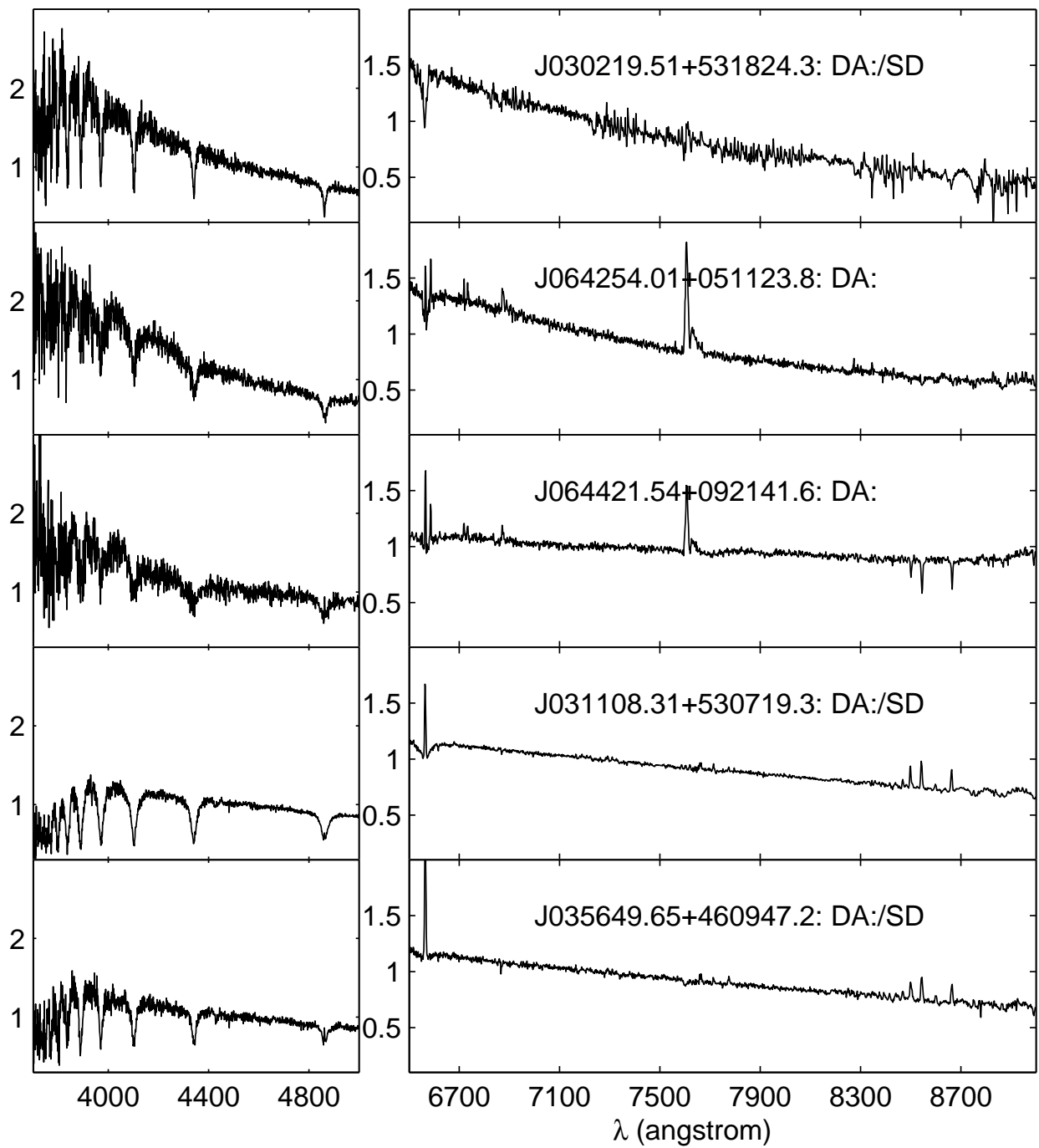}
\figcaption{Examples of spectra with emission lines. the blue part of each spectrum is scaled by the mean flux between 4500 and 4600~\AA~ and the red part is scaled by the mean flux between 7000 and 7500~\AA. The first spectrum is a normal ``DA:/SD" target. The second and the third are ``DA:" targets with emission lines in the center of H$\alpha$. The third spectrum exhibits CaII triplets, but the blue part in more like a DAWD. Considering the unknown process related to the emissions, it is classified as ``DA:". The fourth and the fifth are ``DA:/SD" targets with emission lines in H$\alpha$ and even show CaII triplet emissions.}
\end{figure}

\begin{figure}
\plotone{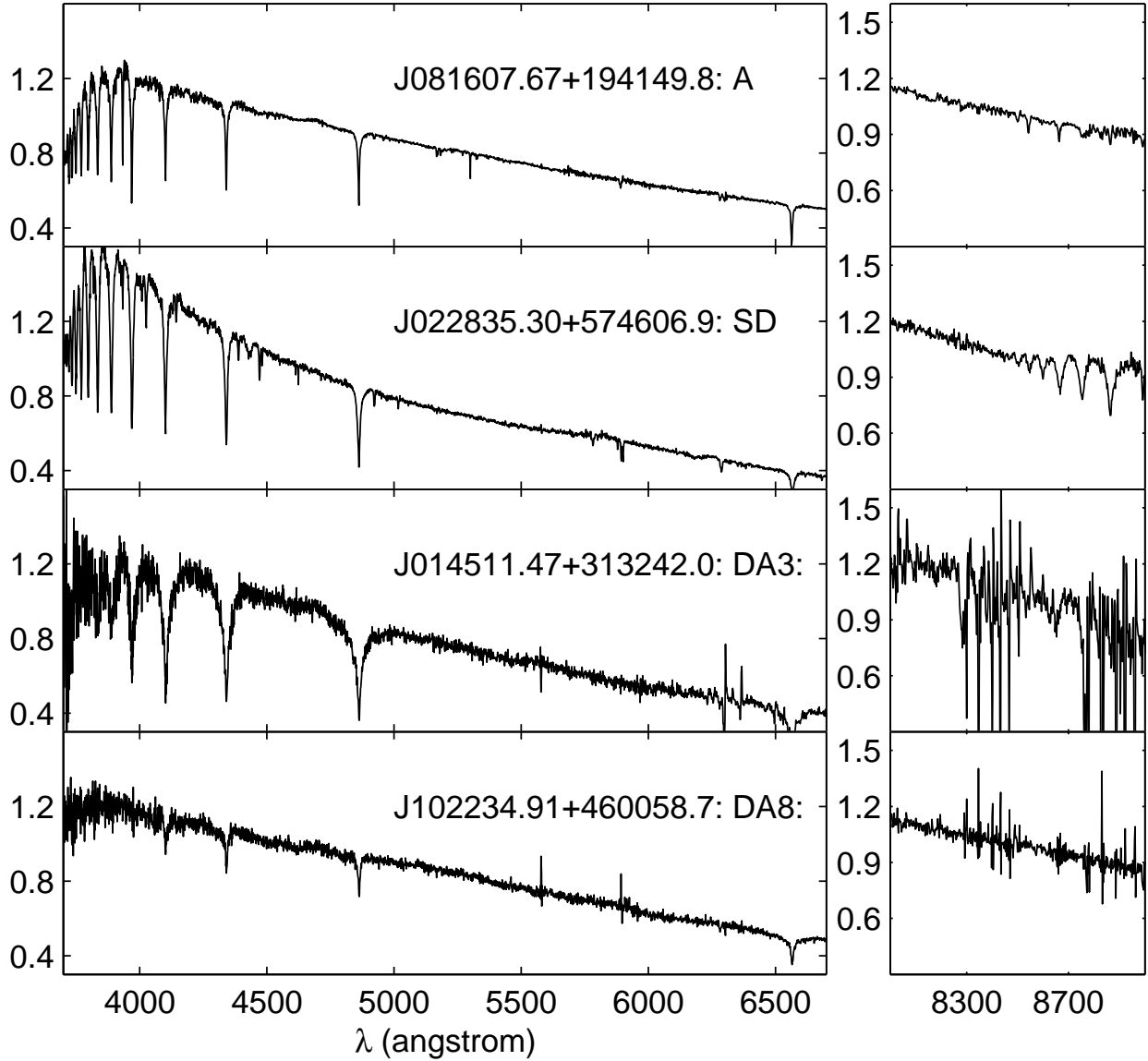}
\figcaption{The first plot is the spectrum of an A-type stars. The second plot is the spectrum of a subdwarf. The third and fourth plot are DAWDs. The left panels show the Balmer series. The spectra are scaled by the mean flux between 4500 and 4600~\AA. The right panels show the wavelength range where CaII triplets and the Paschen lines locate. The spectra are scaled by the mean flux between 8400 and 8600~\AA.}
\end{figure}

\end{document}